\documentclass[twocolumn,preprintnumbers,prd,superscriptaddress,nofootinbib,floatfix,showpacs,showkeys]{revtex4-1}

\usepackage{amsmath}
\usepackage{amsfonts}
\usepackage{amssymb}
\usepackage{graphicx}
\usepackage{color}
\usepackage{hyperref}
\usepackage{booktabs}
\usepackage{hyperref}
\usepackage{cleveref}
\usepackage{braket}
\usepackage{mathtools}
\usepackage[rightcaption]{sidecap}
\usepackage{subfigure}
\usepackage{soul}
\usepackage{enumitem}
\usepackage{comment}

\definecolor{vividviolet}{rgb}{0.62, 0.0, 1.0}
\definecolor{amaranth}{rgb}{0.9, 0.17, 0.31}
\definecolor{palatinateblue}{rgb}{0.15, 0.23, 0.89}
\definecolor{brightpink}{rgb}{1.0, 0.0, 0.5}
\definecolor{cornflowerblue}{rgb}{0.39, 0.58, 0.93}
\definecolor{deepcarminepink}{rgb}{0.94, 0.19, 0.22}
\definecolor{radicalred}{rgb}{1.0, 0.21, 0.37}
\hypersetup{ linktoc=all,
    colorlinks, linkcolor={palatinateblue},
    citecolor={brightpink}, urlcolor={amaranth}
}

\newcommand{\be}{\begin{equation}}
\newcommand{\ee}{\end{equation}}
\newcommand{\bs}{\begin{split}}
\newcommand{\bea}{\begin{eqnarray}}
\newcommand{\eea}{\end{eqnarray}}

\newcommand{\bes}{\begin{subequations}}
\newcommand{\ees}{\end{subequations}}

\begin{document}

\title{A thermodynamic model of inflation without inflaton field}

\author{Jesus Anaya-Galeana}
\email{jesus.anaya@correo.nucleares.unam.mx}
\affiliation{Instituto de Ciencias Nucleares, Universidad Nacional Aut\'onoma de M\'exico, AP 70543, M\'exico, DF 04510, Mexico.}

\author{Orlando Luongo}
\email{orlando.luongo@unicam.it}
\affiliation{School of Science and Technology, University of Camerino, Via Madonna delle Carceri, Camerino, 62032, Italy.}
\affiliation{Istituto Nazionale di Fisica Nucleare (INFN), Sezione di Perugia, Perugia, 06123, Italy.}
\affiliation{SUNY Polytechnic Institute, 13502 Utica, New York, USA.}
\affiliation{INAF - Osservatorio Astronomico di Brera, Milano, Italy.}
\affiliation{Al-Farabi Kazakh National University, Al-Farabi av. 71, 050040 Almaty, Kazakhstan.}

\author{Hernando Quevedo}
\email{quevedo@nucleares.unam.mx}
\affiliation{Instituto de Ciencias Nucleares, Universidad Nacional Aut\'onoma de M\'exico, AP 70543, M\'exico, DF 04510, Mexico.}
\affiliation{Dipartimento di Fisica and Icra, Universit\`a di Roma “La Sapienza”,
Piazzale Aldo Moro 5, I-00185 Roma, Italy.}

\begin{abstract}
In the framework of geometrothermodynamics, we explore how to construct an inflationary cosmic fluid without requiring inflaton fields. To do so, we employ standard radiation and matter as  barotropic background, speculating on the existence of a further real fluid, interpreted as a possible matter with pressure constituent. We thus illustrate its main characteristics, showing how it may face the flatness and horizon problems, yielding an equivalent number of e-foldings, compatible with theoretical expectations. Moreover, we calculate the associated equilibrium and kinetic temperatures, and compute out-of-equilibrium properties, in terms of particle production and effective reheating mechanism. Accordingly, we revise how inflation ends and how to fix the free parameters of our model from observations. Last but not least, we propose how to obtain a bare cosmological constant that could relate the late-time acceleration to inflation, aiming to unify the two approaches. Finally, we discuss the corresponding theoretical implications of our treatment in contrast to the standard approach reliant on scalar field potentials, emphasizing the main advantages and disadvantages.
\end{abstract}



\maketitle
\tableofcontents

\section{Introduction}

Inflation represents a consolidated mechanism driving the universe to speed up at primordial-times \cite{infreview1,2003hep.ph....4257T}. Through a strong accelerated phase, it is possible to heal different drawbacks of the standard Big Bang model, among which the flatness and horizon problems \cite{histinflacion3,caotincinflation}. At the same time, immediately after the inflationary period, baryons are produced during a reheating phase \cite{1994PhRvL..73.3195K}. There, inflationary perturbations are believed to play a crucial role in the clustering of cosmic structures, with remarkable consistency with observational data \cite{infrev2}.

Despite its highly persuasive nature, the inflationary model is guided by the inflaton, an unspecified field whose fundamental characteristics remain elusive \cite{criticsinflation}. Specifically, minimal extensions of the standard model of particle physics could shed light on the inflaton nature, albeit none of the present scenarios appears nowadays really conclusive\footnote{Examples may be even offered by extensions and/or modifications of general relativity (GR), or by including the Higgs fields, spinors, etc., see e.g. Refs. \cite{PhysRevD.32.2511,BEZRUKOV2008703,AntonioDeFelice_2011}. For an in-depth analysis, see e.g. \cite{enciclopedia}.}. In addition, reheating ensures that the temperature increases  immediately after that inflation ends. However, this process is contingent on a specific selection of the driving inflationary model and may conflict with constraints imposed by the Planck satellite on inflationary potentials \cite{plank2018,enciclopedia}.

On the other hand, late-time barotropic fluids, rather than fields, are  consolidated constituents whose use reproduces the features of dust, radiation, dark energy, spatial curvature, and so on \cite{ellis2012relativistic,dodelson2020modern,Perkovi__2020}. Naively, in a homogeneous and isotropic universe, if one takes $\rho\sim a(t)$, with $a(t)$ the cosmic scale factor, there is no barotropic fluid that \emph{simultaneously} can dominate over radiation, providing a sufficient negative pressure to induce inflation at primordial-times. Phrasing it differently, if one takes $\rho\sim a^{-n}$, with $a$ the scale factor and $n$ a natural number, it is impossible to argue any $n$ furnishing a fluid that, during inflation, dominates over radiation, by providing negative pressure.

However, as possible \emph{no-go approach} to construct more complicated thermodynamic fluids that can describe inflation in lieu of scalar fields has been recently developed either under the form of \emph{single fluid scenarios} \cite{singlefluiddarksector, singlefluiddarksector2, Hipólito-Ricaldi_2009} or as unified dark energy-dark matter models \cite{2024PhRvD.109b3510D,2019EPJC...79..332B,Wang_2016, Mishra_2023}. There, the idea is that a single thermodynamic fluid can exhibit a chameleon equation of state that changes according to the universe's expansion and may describe it at different stages\footnote{Additional examples of such scenarios permit to solve the cosmological constant problem \cite{2018PhRvD..98j3520L}, to unify inflationary stage with late-time dynamics \cite{2022CQGra..39s5014D,2024PDU....4401458B,2023CQGra..40j5004B}, predicting dark matter abundance \cite{2021arXiv211205730L} and so on.}.

Consequently, the need for a \emph{covariant versions of thermodynamics}, preserving Legendre invariance while remaining Lorentz invariant, can serve as a crucial foundation for describing inflation without scalar fields\footnote{An illustrative example of this concept can be found in the Fisher-Rao scenario \cite{amari2012differential,Rao1992}, where the geometry within the thermodynamics manifold establishes a specific definition of distance, or metric, which can be employed to define length and probability distributions.} \cite{Orlando1}. In this regard, Geometrothermodynamics (GTD) emerges as a compelling approach that demands the  Legendre invariance at the metric level itself \cite{fundGTD}. This requirement has garnered extensive investigation within the realms of cosmology \cite{2013arXiv1306.6758B,gtdcosmo3,gtdcosmo4,gtdcosmo5}, astrophysics \cite{gtdbh1,gtdbh2,gtdbh3}, and thermodynamics \cite{gtdtermo1,gtdtermo2,gtdtermo3}, yielding promising outcomes, see e.g. \cite{gtdtermo4,gtdapp1,EconoGTD}.

In this study, in the framework of GR, we adopt GTD as the fundamental background to eliminate the need for scalar fields in inflationary realms. Specifically, we demonstrate under which conditions inflation can occur by mimicking the behavior of scalar fields through the use of particular constitutive equations. To achieve this goal, from basic thermodynamic demands, we incorporate into the constitutive equation a matter fluid, a radiation contribution, and an unspecified real fluid. The inclusion of a real fluid is motivated by the general expectation that real fluids are more prevalent than perfectly approximated ones. In other words, we employ genuine barotropic fluids plus an additional one, whose fundamental properties can be used to frame the universe as a thermodynamic system. Thus, we show that it is possible to replicate the inflationary dynamics with no need for scalar particles, but directly through the inclusion of the underlying mixture of fluids. Significantly, we choose our free constants resembling the rapid expansion observed during a quasi-de Sitter phase and require our effective pressure for radiation and matter to sub-dominate compared with our real fluid. We categorize the origin of inflation as arising from a phase transition, a concept naturally accommodated within the framework of real fluids. Afterwards, we explore how inflation ends and the conditions necessary for a possible reheating mechanism to take place. Importantly, we demonstrate that akin to the inflationary phase, the late-time accelerated expansion of the universe can naturally occur, reconciling \emph{de facto} inflation and dark energy. Finally, in all our treatment, we compare our overarching GTD fluid with existing theoretical predictions and modern observational data.

The paper is structured as follows. In Sect. \ref{sezione2}, the basic demands of inflation have been reported, while in Sect. \ref{sezione3}, we illustrate how to construct an inflationary thermodynamic cosmic fluid, without including the scalar field inflaton. We also show how to obtain inflation through this fluid, focusing on the fact that the fluid might overcome radiation and matter throughout the period in which the inflationary acceleration occurs. Afterwards, in Sect. \ref{sezione4}, the Big Bang inconsistencies are faced through our fluid, giving emphasis on how dark energy can emerge from the same unified picture. Finally, Sect. \ref{sezione5} is devoted to the conclusion and perspectives of our effort, remarking that our scenario represents an alternative to standard inflation made by cosmic fluids obtained from thermodynamics, in a unified picture in which inflation and dark energy are obtained simultaneously.


\section{Inflationary setup}\label{sezione2}

In this section, we report the main hypotheses made to guarantee an inflationary epoch. As the first assumption, we undertake the
Friedmann-Lema\^itre-Robertson-Walker metric
\begin{equation}
    \label{eq: FLRW metric}
    ds^2 = - dt^2 + a^2(t)\left(\frac{dr^2}{1 - kr^2} + r^2d\theta^2 + r^2\sin^2 \theta d\phi^2 \right) .
\end{equation}
where $t$ is the cosmic time and $(r, \theta ,\phi)$ are the usual spherical coordinates. Particularly, Eq. \eqref{eq: FLRW metric} involves the topological spatial curvature, $k$, and the scale factor, $a(t)$. Hereafter, if necessary we will assume the universe as spatially flat. Solving the Einstein equations with Eq. (\ref{eq: FLRW metric}) implies to write up the Friedmann equations:
\begin{subequations}
    \begin{align}
    \label{eq: friedmann1}
    &\frac{\dot{a}^2}{a^2}\equiv H^2 = \frac{8 \pi G_N}{3}\rho  - \frac{k}{a^2}.\\
    \label{eq: friedmann2}
    &\frac{\ddot{a}}{a} = \dot H+H^2=- \frac{4\pi G_N}{3}\left(\rho + 3P \right),
    \end{align}
\end{subequations}
and, alternatively, as a combination of the above equations,  the continuity equation,
\begin{equation}
    \label{eq: continuity}
   \dot{\rho} + 3\frac{\dot{a}}{a}\left( \rho + P \right) = 0 .
\end{equation}

If $P$ and $\rho$ are due to a scalar field, $\phi$, we could write the corresponding Lagrangian, that formally reads
\begin{align}
\label{eq:1}
\mathcal{L}^Q &= X -V\left(\phi\right)\,,
\end{align}
where $X \equiv\frac12 g^{\alpha\beta}\partial_\alpha \phi \partial_\beta\phi$ is the kinetic term and $V\equiv V(\phi)$ is the potential that drives inflation, having labelled with $Q$ the Lagrangian to indicate the general concept of \emph{quintessence}.

From Eq. \eqref{eq:1}, we easily obtain
\begin{subequations}\label{eq:no11}
\begin{align}
\rho =\, &2X \mathcal{L}_{,X} - \left(X -  V \right)=\frac{1}{2}\dot \phi^2+V\,,\\
\label{eq:no12}
P =\, &X - V=\frac{1}{2}\dot \phi^2-V\,.
\end{align}
\end{subequations}
where the last passages adopted the homogeneity and isotropy of the universe to simplify the density and pressure.

\subsection{Including slow roll}

Considering dots and primes as time and field derivatives respectively, the inflationary period undergoes the  \emph{slow roll} stage, where the potential energy of the inflaton dominates over its kinetic energy, and
\begin{equation}
\dot\phi^2\ll V(\phi)\>\Rightarrow\>\ddot\phi\ll H\dot\phi,\label{2 slowroll}
\end{equation}
implying
\begin{align}
    H^2&\simeq\frac{8\pi}{3M_{Pl}^2} V(\phi),\label{srfr}\\
    3H&\dot\phi\simeq-V^\prime(\phi),\label{srdin}
\end{align}
encoded in the \emph{slow roll parameters} defined as
\begin{subequations}
\begin{align}
    \epsilon_H(\phi)&\equiv\frac{M_{Pl}^2}{4\pi\mathcal{L}_{,X}}\left(\frac{H^\prime(\phi)}{H(\phi)}\right)^2,\label{heps}\\
    \eta_H(\phi)&\equiv\frac{M_{Pl}^2}{4\pi\mathcal{L}_{,X}}\frac{H^{\prime\prime}(\phi)}{H(\phi)}.\label{heta}
\end{align}
\end{subequations}
sometimes rewritten as
\begin{subequations}
\begin{align}
    \epsilon_V&\equiv-\frac{\dot H}{H^2}\simeq\frac{M_{Pl}^2}{16\pi\mathcal{L}_{,X}}\left(\frac{ V^\prime}{ V}\right)^2,\label{eps}\\
    \eta_V&\equiv-\frac{\ddot \phi}{H\dot\phi} -\frac{\dot H}{H^2} \simeq\frac{M_{Pl}}{8\pi}\left(\frac{V^{\prime\prime}}{ V}\right).\label{eta}
\end{align}
\end{subequations}
that provide $H^2\gg|\dot H|\simeq0$, giving rise to a quasi-de Sitter phase  \cite{histinflacion4,linde2005particle}.

To sustain inflation, it is necessary then a further slow roll condition imposed by hands. From these considerations, we require
\begin{equation}
    \ddot{a}>0\>\iff\>\epsilon_H\ll1,
\end{equation}
and so, inflation ends as $\epsilon_{V;H}\simeq 1$.

The initial and final inflaton values are $\phi_i$ and $\phi_f$, respectively, and so to guarantee inflation, the total number of e-foldings is usually calculated as
\begin{equation}
    N\equiv\int_{\tau_i}^{\tau_f}{Hd\tau}\simeq-\frac{8\pi}{M_{Pl}^2}\int_{\phi_i}^{\phi_f}{\frac{ V}{ V^\prime}d\phi},\label{efold slowroll1}
\end{equation}
requiring  $N\gtrsim60$, as in the standard inflationary picture \cite{efoldslimit1,efoldslimit2}.

Summarizing, one needs to first understand the form of the underlying inflationary potential and, additionally, two more  assumptions are required: the mechanism through which inflation ends (commonly accepted as a graceful exit determined from a chaotic potential) and the slow roll phase, which ensures the strong acceleration period.

Consequently, inflation appears more complicated than simply assuming a real fluid driving the universe dynamics. While the chaotic nature of inflationary potentials is widely accepted to explain the end of inflation, it remains challenging to determine whether the inflationary potential is modelled through small or large fields and how such a potential can be reconstructed.

\section{Thermodynamic  inflation}\label{sezione3}

Inflation is commonly referred to as a strong accelerated phase that occurs once the scalar fields obey the slow roll approximation. Nevertheless, as stated above, standard versions of barotropic fluids cannot account for a complete description of inflationary time.

Alternatively, we seek new versions of these fluids that show up viable constitutive equations, able to speed the universe up.

For example, it was previously shown that,  involving the entropy $S$, the energy and volume, $U$ and $V$,  respectively, and following the technique reported in Ref.  \cite{CosmoGTD}, a possible fundamental relation of the form
\begin{equation}
    \label{eq: fund equation}
     S(U,V) = c_1 \ln \left(U+\frac{\alpha }{V}\right)+c_2 \ln (V-\beta ),
\end{equation}
allows to derive the standard barotropic equation used in the standard $\Lambda$CDM paradigm when $\alpha=\beta=0$ i.e $P = \omega \rho$, providing a constraint over the free parameters,  $\omega = c_2/c_1$.

The fundamental equation, written in GTD, depends on two free parameters, $\alpha$ and $\beta$, interpreted analogously to those related to the Van der Waals fluid in standard thermodynamics. Specifically, $\alpha$ is associated with thermodynamic interactions, while $\beta$ represents the volume occupied by the particles constituting the fluid\footnote{It is worth noting that there exists a maximum value for each of these parameters. Precisely, $\alpha/V$ cannot exceed the total energy content of the universe itself, as it exhibits the units of energy, whereas, similarly, the maximum value for $\beta$ is constrained by the universe volume.}.

In analogy to the $\Lambda$CDM case, we wonder whether a constitutive equation may exist to describe inflationary stages. Hence, the above constitutive equation may serve as a prototype for establishing the minimal conditions required to ensure the onset of inflation, as we report below, i.e., without the need of a further scalar field, playing the role of inflaton.

\subsection{Conditions}

Hence, to show how the cosmic speed up can arise from our GTD fluid, we recall that - in a spatially flat universe - our energy density, $\rho$, behaves approximately as a constant quasi-de Sitter energy density, so that $\epsilon_H \ll 1$ with
\begin{equation}
    \label{eq: def hubble slow-roll param}
    \epsilon_{H} \equiv -\frac{d \ln H}{d \ln a} = - \frac{\Dot{H}}{H^2},
\end{equation}
turning into

\begin{equation}
    \label{eq: def hubble slow-roll param en densidad}
    \epsilon_{H} = 1 + \frac{4\pi G}{3}\frac{1}{H^2}\left(\rho + 3P\right).
\end{equation}

This is equivalent to demand that $P/\rho = -1 + \Delta$, where $P$ is the pressure, $\rho$ the energy density and $\Delta$ is such that $|\Delta| \ll 1$. In general, it can be said that this last condition is necessary for inflation to hold.

Let us return to Eq. (\ref{eq: fund equation}): There, by calculating the equations of state and introducing the following notations $U/V = \rho$, $V = V_0 (a/a_0)^3$, we can write $P=P(\rho, a)$; this combined with the continuity equation, Eq. (\ref{eq: continuity}), gives us the energy density as a function of the scale factor:
\begin{equation}
    \label{eq: density inf fluid}
    \rho = K \frac{ \left( a^3V_0 - \beta \right)^{-\frac{c_2}{c_1}}}{a^3} - \frac{\alpha}{a^6V_0^2} \,,
\end{equation}
where $K$ is an integration constant. Eq. (\ref{eq: density inf fluid}) is roughly a constant as
\begin{enumerate}
    \item the first term dominates, namely $\alpha \ll V_0^2a_i^6\rho_i$,
    \item if $c_2/c_1 \approx - 1$ and,
    \item as $\beta \lesssim V_0{a_i}^3$.
\end{enumerate}
Above, $V_0$ is the volume of the universe today, $a_i$ is the scale factor at the beginning of inflation, and $\rho_i$ is the energy density at the beginning of inflation.
The first and last conditions can be written in a different form,  by introducing a new parameter $\chi$ that relates the value of both $\alpha$ and $\beta$ to their maximum (at the beginning of inflation):
\begin{subequations}
\label{eq: chi}
    \begin{align}
    &\chi = \frac{\beta}{\beta_c} = \frac{|\alpha|}{\alpha_c} ,  \\
&\alpha_{max} := V_0^2a_i^6\rho_i  ,\\
&\beta_{max} := V_0a_i^3 .
    \end{align}
\end{subequations}
Therefore, for $\rho$ to be a (quasi) constant, we need $\alpha<\alpha_{max}$, $\beta<\beta_{max}$ or $\chi<1$.
Now the equation of state that relates the pressure $P$ and scale factor $a$ turns into:
\begin{equation}
    \label{eq: EOS inf fluid}
    P = \frac{c_2}{c_1}\rho - \frac{\alpha}{a^6{V_0}^2}\left(1-\frac{c_2}{c_1}\right) .
\end{equation}

If $\chi<1$ and $c_2/c_1 \approx - 1$, then we get $P/\rho = -1 + \Delta$ and we end up with a quasi-De Sitter phase.
Roughly, selecting $c_2/c_1 = -8/9$ we get a number of e-foldings $N$ equal to
\begin{equation}
    N\simeq 56,
\end{equation}
which, combined with scales at which we believe Grand Unification theories may occur, i.e., at energies  $\sim 10^{15} \, \, \, GeV$ at which inflation can take place \cite{inflationenergyscale,inflationenergyscale2}, sets the beginning of inflation at
\begin{equation}
a_i = 8.75 \cdot 10^{-54}\,.
\end{equation}

\subsection{Phase transition}

The current understanding of inflation can be divided into models underlying a phase transition, related to the so-called \emph{old inflation} and \emph{chaotic models}, where the graceful exit avoids the need of a phase transition.

We thus wonder whether our approach requires additional properties to enable inflation to start.

Our standpoint is to work a real fluid out and, following the above example, assuming a Van der Waals fluid, guaranteeing a phase of strong acceleration.

It is widely-known that a Van der Waals fluid can exhibit a thermodynamic phase transition, which is usually seen in the Gibbs potential, or the isotherm curves in a $P-V$ plot \cite{Callen}.

To develop this specific case, the approach with different levels of temperature does not appear particularly interesting, since the hypothesis of thermodynamic equilibrium is far from being reached in our picture due to the fact that the temperature of this system is a function of the scale factor itself and, in fact, we end up with
\begin{equation}
    \label{eq: temperature inf fluid}
    T = \frac{V_0a^3}{c_1}\left(\rho + \frac{\alpha}{a^6{V_0}^2}\right).
\end{equation}
To develop a self-consistent approach, we investigate the curvature singularities of the equilibrium phase space of our system that may denote the existence of phase transitions in curved manifolds, see e.g. \cite{fundGTD, PhaseGTD,VdWphasetransGTD} that, for a Van der Waals fluid, will denote the scale factor at which the phase transition occurs.

In virtue of this, \emph{we can interpret thermodynamic phase transition as the process that gave rise to the inflationary epoch.} Indeed, let us focus on the term $\alpha/V$, i.e., the one that generates phase transitions \cite{Johnston_2014}. Since the volume is proportional to $a^3$, this term tends to zero as the universe expands and, so, this implies that the Van der Waals fluid tends to be ideal as the universe expands, i.e., as the universe gets bigger, the fluid dilutes, leading to an end for the inflationary period since we have:
\begin{equation}
    \epsilon_{H} \approx \frac{3}{2}\left[ \frac{1}{1+\frac{c_2}{c_1}} - \frac{1}{1+\frac{c_2}{c_1}}\chi\left(\frac{a_i}{a}\right)^3 -2\chi\left(\frac{a_i}{a}\right)^{-3\left(\frac{c_2}{c_1}-1\right)} \right]^{-1}
\end{equation}

For large enough $a$, the condition $\epsilon_H \ll 1$ breaks down, having $\epsilon_H \simeq 1$.

Hence, in our picture we invoke inflation to naturally end, as the fluid diluites.


\subsection{Reheating}

We interpret the effective constituent of this Van der Waals-like fluid as \emph{ matter with pressure}, at primordial-times. In the epoch immediately after the Big Bang, this assumption is clearly possible since the fluid lies at equilibrium at very high temperature.

Nevertheless, by itself the model cannot account for reheating, as the net fluid is one. We thus propose that the fundamental equation is slightly modified by the following assumptions:
\begin{itemize}
    \item[-] Dynamics induced by perfect fluids drives the universe to pass through matter and radiation- dominated phases.
    \item[-] Real fluids, or more broadly non-barotropic fluids, are responsible for universe phase transition that can induce phases of strong acceleration.
\end{itemize}
In other words, the entropy, namely the fundamental equation can be split into
$S_T(U,V)=S_{ideal}+S_{non-ideal}$,
where $S_{ideal}$ contains all the perfect fluids that do not imply accelerated phases, while $S_{non-ideal}$ is built up by those fluids that may induce a phase transition, namely a phase of strong acceleration, under certain circumstances, that can be interpreted as inflation.

Following this recipe, we can easily propose
\begin{equation}
S_T(U,V)=S_{matter}+S_{radiation}+S_{non-ideal}\,.
\end{equation}
Indeed, the entropy is given by an unspecified sum over the possible species of ideal and non-ideal fluids, say $S=\sum_i S_i=S_1+S_2+S_3+\ldots$.

As a possible minimal choice for the fundamental equation, we thus propose
\begin{equation}
    \label{eq: fund equation 3 fluids}
    \begin{split}
     S_T(U,V) =& S_{inf}+ S_{rad} + S_{dust} \\
     =& c_1 \ln \left(U+\frac{\alpha }{V}\right)+c_2 \ln (V-\beta ) \\
     &+ c_3 \ln (U)+c_4 \ln (V)  \\
     & + c_5\ln (U) ,
    \end{split}
\end{equation}
where we specified $S_{inf}$ as in Eq. \eqref{eq: fund equation}. The sets of constants, $c_i, i=1;5$ are arbitrary and have not been fixed at this stage.

Hence, from the first law of thermodynamics, $TdS = dU + PdV$, we have
\begin{equation}
    \label{eq: 1/t inflationary epoch}
    \frac{1}{T} = \frac{\partial S}{\partial U} = \frac{c_1 V}{\alpha +U V}+\frac{3 c_4 + c_5}{U} ,
\end{equation}
\begin{equation}
    \label{eq: p/t inflationary epoch}
    \frac{P}{T} = \frac{\partial S}{\partial V}= -\frac{\alpha  c_1}{U V^2+\alpha  V}+\frac{c_2}{V-\beta }+\frac{c_4}{V} .
\end{equation}

Recalling again, $U/V = \rho$ and $V = V_0 a^3$, we obtain
\begin{widetext}
\begin{equation}
    \label{eq: Pressure inflationary epoch}
    P(\rho,a)=\frac{\rho}{V_0a^3 -\beta}\left[\frac{-\alpha c_1 \left(V_0a^3 -\beta \right) + c_2 V_0a^3 \left(\rho {V_0}^2a^6+\alpha\right) + c_4\left(V_0a^3 -\beta\right)\left(\rho {V_0}^2a^6+\alpha\right)}{c_1\rho{V_0}^2a^6+(3c_4+c_5)\left(\rho {V_0}^2a^6+\alpha\right)}\right],
\end{equation}
\end{widetext}
that is equivalent to the algebraic relation, once using Eq. (\ref{eq: continuity}),
\begin{equation}
    \label{eq: algebraic cont equation}
\begin{split}
    \left[a^6 \rho (a)\right]^{3 c_4+c_5} & \left[ \alpha +  a^6 V_0^2 \rho (a) \right]^{c_1} = \\
    &\xi a^{3 (c_1+2 c_4+c_5)} \left(V_0a^3 -\beta\right)^{-c_2},
\end{split}
\end{equation}
with $\xi$ an integration constant.

Since $\alpha<\alpha_{max}$, $\beta<\beta_{max}$ or equivalently $\chi<1$, it follows that Eq. (\ref{eq: algebraic cont equation}) takes the form
\begin{equation}\label{eq: rho(a) segunda aprox}
     a^6\rho (a) \approx \xi {V_0}^{-\frac{c_2+2c_1}{3c_4+c_5+c_1}}a^{\nu_1 },
\end{equation}
with $
\nu_1  \equiv \frac{3 (c_1 + 2c_4 + c_5 - c_2)}{3c_4+c_5+c_1}$.

Immediately, from Eq. (\ref{eq: rho(a) segunda aprox}), we can see the condition needed to flatten $\rho$ to a constant, namely
$\nu_1 = 6 $, which is satisfied if we assume:
\begin{enumerate}
    \item[-] $\alpha < \alpha_{max}$,  $\beta  < \beta_{max}$ or equivalently $\chi<1$
    \item[-] $c_2 =  -4c_4 - c_1 - c_5$.
\end{enumerate}

Then, the energy density becomes
\begin{equation}
    \label{eq: rho(a) inflationary epoch}
    \begin{split}
    \rho (a) &\approx \xi {V_0}^{\frac{4c_4+c_5-c_1}{3c_4+c_5+c_1}}  \\
    \end{split}
\end{equation}
while the pressure reads
\begin{equation}
    \label{eq: P(a) inflationary epoch}
    \begin{split}
    P(a) &=  -\left[1 + \frac{\beta}{V_0a^3}\left( \frac{4c_4+c_5+c_1}{3c_4+c_5+c_1}\right)\right]\rho \\
    &\, \, \, \, \, - \left(\frac{3c_4+c_5+2c_1}{3c_4+c_5+c_1}\right)\frac{\alpha}{{V_0}^2a^6} \\
    & \approx - \rho,
    \end{split}
\end{equation}
probing that Eq. \eqref{eq: fund equation 3 fluids} exhibits an inflationary epoch.

However, inflation will continue for evern unless  particles are produced at the end of the process, as we point out in the next subsection.

\subsection{Particle production}

To stop inflation and, consequently, to include the possibility of having particle production, we can check what happens using the Boltzmann equation.

In the usual scalar field, the reheating phase is accounted by coupling the inflaton with the environment field, so that the energy stored in the inflaton field can be used to create new relativistic particles

Following this idea, we wonder whether this can happen even without scalar field. Hence, by introducing three thermodynamic variables,

\begin{itemize}
    \item[-] the particle number for the inflaton fluid, hereafter referring to as inflationary particle number,
    \item[-] the particle number for the radiation fluid, namely radiation particle number, and
    \item[-] the particle number for the dust fluid, say dust particle number.
\end{itemize}

In this scheme, as our fluid is not ideal, we will interpret the reheating phase by allowing the inflationary particles to decay into relativistic and dust particles\footnote{Clearly, particle generation in cosmology is the underlying assumption behind this choice. Mechanisms for particle creation are well known in the literature, see e.g. \cite{Ford:2021syk}, with promising results provided during inflationary stages, as shown in Refs. \cite{Belfiglio:2024xqt,Belfiglio:2022yvs}.}.

The consequent process, namely allowing particles to decay, appears related to non-equilibrium, as a consequence of the universe expansion.

Reaching the subsequent equilibrium will therefore stop inflation, being responsible for the reheating phase. There, if a particle X decays into a particle Y and a particle Z, then the Boltzmann equations are \cite{primordialcosmology}
\begin{subequations}
    \label{eq: Boltzmann equation decay}
    \begin{align}
        \frac{\partial f_X}{\partial t} - \frac{\Dot{a}}{a}p\frac{\partial f_X}{\partial p} =& - \frac{f_X}{\tau_X} , \\
        \frac{\partial f_Y}{\partial t} - \frac{\Dot{a}}{a}p\frac{\partial f_Y}{\partial p} =& - \lambda \frac{f_X}{\tau_X} , \\
        \frac{\partial f_Z}{\partial t} - \frac{\Dot{a}}{a}p\frac{\partial f_Z}{\partial p} =& - ( 1 -\lambda)\frac{f_X}{\tau_X} ,
    \end{align}
\end{subequations}
where $f_{i}$ is the distribution function for each species, $\tau_{X}$ the mean lifetime of the $X$ particle, $\lambda$ a dimensionless parameter that controls the number of $Y$ particles that are created, and $t$ is the cosmic time.

From the definition of the particle number density
\be
n_A = \frac{g_{dof}}{(2\pi)^3}/\int d^3p f_i = \frac{N_A}{V},
\ee
we have:
\begin{subequations}
    \label{eq: number density equation decay}
    \begin{align}
        &\frac{d n_X}{\partial t} - \frac{\Dot{a}}{a}n_x + \frac{n_X}{\tau_X} =  0, \\
        &\frac{d n_Y}{\partial t} - \frac{\Dot{a}}{a}n_Y - \frac{\lambda n_X}{\tau_X} =  0, \\
        &\frac{d n_Z}{\partial t} - \frac{\Dot{a}}{a}n_Z - \frac{(1-\lambda)n_X}{\tau_X} =  0.
    \end{align}
\end{subequations}
The solution of this system can be expressed as
\begin{subequations}
    \label{eq: number density}
    \begin{align}
        n_X & \propto \frac{1}{a^3}e^{-t/\tau_X}. \\
        n_Y &\propto\frac{ \lambda}{a^3}\left[1- e^{-t/\tau_X}\right]. \\
        n_Z &\propto \frac{(1-\lambda)}{a^3}\left[1-e^{-t/\tau_X}\right].
    \end{align}
\end{subequations}
As $V \propto a^3$, the particle numbers become:
\begin{subequations}
    \label{eq: particle number}
    \begin{align}
        N_X & \propto e^{-t/\tau_X}, \\
        N_Y &\propto \lambda \left[1-e^{-t/\tau_X}\right], \\
        N_Z &\propto (1-\lambda)\left[1-e^{-t/\tau_X}\right].
    \end{align}
\end{subequations}

Without losing generality, we now require that $X$ represents the inflationary particles, whereas $Y$ and $Z$ represent the radiation and dust particles, respectively.

Now, if we further suppose that there exists an inflationary fluid that does not decay, i.e.,  we have $\lambda \in \left[0,\zeta\right]$, $\zeta \in [0,1]$ ($\zeta \ll 1$, so that radiation completely dominates after inflation) a  dimensionless parameter that controls the amount of inflationary fluid remaining after the decay.

We must have (considering that the particle number is conserved)
\begin{subequations}
    \label{eq: particle number conserved}
    \begin{align}
        N_{inf}  &= N_i(1 - \zeta) e^{-t/\tau_{inf}} + N_i\zeta, \\
        N_{rad} &= \lambda(Ni -  N_{inf}), \\
        N_{dust} &= (1-\lambda)(Ni -  N_{inf}).
    \end{align}
\end{subequations}
where $N_i$ the total number of particles.

Finally, to seek the dependence in terms of the scale factor, in an Einstein-de Sitter universe, dominated by radiation, we have   $t\propto a^2$ and, since after inflation radiation dominates, it is plausible to use this approximation to write
\begin{equation}
    \label{eq: numero part inflacion}
    N_{inf}, \simeq  \begin{cases} N_i & a < a_{trans}, \\

    N_i(1-\zeta) e^{-\frac{1}{2\nu_2 }\left[\left(\frac{a}{a_{trans}}\right)^2 - 1 \right]} + N_i\zeta,  & a \ge a_{trans}, \end{cases}
\end{equation}
so that the associated fluid, during inflation, behaves in analogy to the inflaton, while right after some scale factor, $a_{trans}$, the decay will start, letting radiation and dust to dominate after the inflationary stage\footnote{In the above equations, we have defined $\tau_{inf} = \nu_2   \, a_{trans} \sqrt{\frac{3}{8\pi G \rho_{inf}}}$; $\nu_2 $ a non-dimensional constant ($\nu_2  \in \mathbb{R}^{+}$) and $a_{trans}$ the scale factor when the transition starts.}.

However, accounting for a species decay into others is not well-provided by the previous fundamental equation, Eq. (\ref{eq: fund equation 3 fluids}).

Hence, the introduction of the three new thermodynamic variables modifies inevitably the fundamental equation, that turns into
\begin{widetext}
\begin{equation}
    \label{eq: ec fund con particulas}
    \begin{split}
     S(U,V) &= c_1N_{inf} \ln \left(\frac{U}{N_{inf}}+\frac{\alpha N_{inf} }{V}\right)+c_2N_{inf}\ln \left(\frac{V}{N_{inf} }-\beta \right)  + 3c_4N_{rad}  \ln \left(\frac{U}{N_{rad}}\right)  \\ & \, \, \, \, \, +c_4N_{rad} \ln \left(\frac{V}{N_{rad}}\right) + c_5N_{dust}  \ln \left(\frac{U}{N_{dust}}\right).
    \end{split}
\end{equation}
\end{widetext}

For this system, the pressure changes accordingly,
\begin{widetext}
\begin{equation}
    \label{eq: Presion con num de part}
    \begin{split}
    P(\rho,a) =& \frac{\rho}{V_0a^3 -N_{inf}\beta}\left[\frac{1}{N_{inf}c_1\rho{V_0}^2a^6+\left(3N_{rad}c_4+N_{dust}c_5\right)\left(\rho {V_0}^2a^6+N_{inf}^2\alpha\right)}\right] \\
    &\left[-\alpha c_1 N_{inf}^3\left(V_0a^3 -N_{inf}\beta \right) + c_2 V_0a^3N_{inf} \left(\rho {V_0}^2a^6+N_{inf}^2\alpha\right) +  c_4N_{rad}\left(V_0a^3 -N_{inf}\beta\right)\left(\rho {V_0}^2a^6+N_{inf}^2\alpha\right) \right]  .
    \end{split}
\end{equation}
\end{widetext}

In analogy to the case developed before, we need to solve the continuity equation, Eq. \eqref{eq: continuity}. It can be solved only numerically, as due to the complexity of Eq. (\ref{eq: Presion con num de part}), where we employ the free parameters, $c_1$, $c_2$, $c_4$, $\alpha$, $\beta$, $a_i$, $\rho_i$, $N_i$,$V_0$,$\lambda$,$\zeta$, that are not fixed \emph{a priori}.

To infer their values, in order to numerically implement the equation of state, we notice that the first three parameters can be matched with the energy conditions, whereas the remaining parameters can be easily fixed, as it will be clearer later in the text.

\section{Fixing the free parameters}

To bound the free parameters of the model, we split our treatment in terms of different regimes. To do so, we can single out each epoch and thermodynamic quantity of interest as reported above.


\subsubsection{Initial pressure}

At the beginning of inflation, we need to have $N_i = N_{inf}$, $N_{rad} = N_{dust} = 0$, $P \approx - \rho$.
Recalling the assumptions made initially, we can  include the particle number.

Essentially, making the following  change

\begin{equation}
\alpha \longrightarrow N_i^2 \alpha,\hspace{1cm}\beta \longrightarrow N_i \beta,
\end{equation}
the new conditions become
\begin{equation}
    \label{eq: condiciones con num de part}
    \begin{split}
         N_i^2\alpha  <\alpha_{max} &\hspace{0.5cm} \text{and} \hspace{0.5cm} N_i\beta  < \beta_{max} .
    \end{split}
\end{equation}

Inserting these relations into Eq. (\ref{eq: Presion con num de part}) at the beginning of inflation, say at $a=a_i$, we obtain
\begin{equation}
    \label{eq: presion num part inicio inflacion}
    \left. P \right|_{a_i} \approx  \left. \frac{c_2}{c_1}\rho \right|_{a_i} - \left. \frac{N_i^2\alpha}{{V_0}^2a^6} \right|_{a_i},
\end{equation}
that represents the requested initial pressure.

It is the easy to see that if we select $c_2 = - c_1$, then $P \approx - \rho$ during inflation. The physical interpretations of $\alpha$, and $\beta$ remain unaltered.

\subsubsection{Conditions as inflation ends}

When inflation ends, the particle number obeys the constraint, $N_{rad}>N_{inf}$, i.e., turning out to be a necessary condition to have the pressure $\sim \rho/3$. This can fully-happen when radiation completely dominates, say at $a=a_{rad}$.

Since the inflationary particle number is either a constant or a decreasing function, then Eqs. (\ref{eq: condiciones con num de part}) imply

\begin{subequations}
\begin{align}
&N_{inf}^2\alpha \le N_i^2\alpha < \alpha_{max},\\
&N_{inf}\beta  \le N_i\beta < \beta_{max}.
\end{align}
\end{subequations}

If we also assume that $ \left. 3N_{rad}c_4 \right|_{a_{rad}} + \left. N_{dust}c_4\right|_{a_{rad}} \gg \left. N_{inf}c_1 \right|_{a_{rad}}$,  then the second term of Eq. (\ref{eq: Presion con num de part}) can be written as
\begin{widetext}
$$\frac{1}{N_{inf}c_1\rho{V_0}^2a^6+3N_{rad}c_4\left(\rho {V_0}^2a^6+N_{inf}^2\alpha\right)} \approx \frac{1}{(3N_{rad}c_4+N_{dust}c_5\left(\rho {V_0}^2a^6+N_{inf}^2\alpha\right)}\left(1-\frac{N_{inf}c_1}{3N_{rad}c_4+N_{dust}c_5}\right) \, . $$
\end{widetext}

This ultimately leads to
\begin{equation}
    \label{eq: presion num part final inflation 1 aprox}
    \begin{split}
    \left. P \right|_{a_{rad}}&   \approx \left. \left( 3+\frac{N_{dust}c_5}{N_{rad}c_4}\right)^{-1} \rho \right|_{a_{rad}}  \\
    & \hspace{-0.6cm} -\left. \frac{N_{inf}c_1}{3N_{rad}c_4+N_{dust}c_5} \left[1 + \left( 3+\frac{N_{dust}c_5}{N_{rad}c_4}\right)^{-1} \right] \rho \right|_{a_{rad}} \, .
    \end{split}
\end{equation}

Let us now turn into the physical meaning behind the condition $\left. 3N_{rad}c_4 \right|_{a_{rad}} \left. N_{dust}c_5 \right|_{a_{rad}}  \gg \left. N_{inf}c_1 \right|_{a_{rad}} $, that can be argued starting from the fundamental equation, Eq. (\ref{eq: fund equation}), where the constant, $c_1$, is related to the heat capacity at constant volume per number of particles.

Conversely, in Eq. (\ref{eq: ec fund con particulas}), $c_1$, $c_3=3c_4$ and $c_5$ are the heat capacities at constant volume of the inflationary, radiation and dust fluids, respectively.

Hence, bearing these conditions in mind, we require that the heat capacity at the
radiation-dominated era is given by \emph{radiation and dust}\footnote{Note that this is a sufficient condition for radiation (and dust) to dominate whereas $\left. N_{rad}\right|_{a_{rad}} + \left. N_{dust}\right|_{a_{rad}} \gg \left. N_{inf}\right|_{a_{rad}}$ is just a necessary condition.}.

Afterwards, to infer $P \approx \rho/3$ we still need to note that $N_{dust}c_5 \ll N_{rad}c_4$ suggesting that the heat capacity, in the radiation-dominated era must be (mostly) given by the radiation fluid, certifying the condition on the pressure.

In this case, the necessary condition to let radiation dominate is $\left. N_{rad} \right|_{a_{rad}} \gg  \left. N_{dust} \right|_{a_{rad}} $ whereas $\left. N_{dust}c_5 \right|_{a_{rad}}  \ll \left. N_{rad}c_4 \right|_{a_{rad}} $ is the sufficient one.

These last two requirements allow us to impose constrictions on $\lambda$. The necessary condition imposes $\lambda>0.5$, while the sufficient  $(1-\lambda) c_5\ll c_4 \lambda$.

It is important to note that the conditions on  the heat capacities, i.e., on the constants $c_1$, $c_4$, and $c_5$, indicate the scale factor at which the radiation fluid dominates, and their violation gives information about the end of the transition/decaying process.

\subsection{Conditions on the temperature}

The thermodynamic temperature for this system is
\begin{equation}
    \label{eq: Temp con num de part}
    T(\rho,a) = \left[
    \frac{V_0a^3c_1N_{inf}}{\rho {V_0}^2a^6+N_{inf}^2\alpha}
    + \frac{3c_4N_{rad}+N_{dust}c_5}{\rho V_0a^3} \right]^{-1}  \, .
\end{equation}

After the decaying process, the fluid is clearly mainly made of particles, whose distributions could be either bosonic or fermionic. As the radiation-dominated epoch starts, the corresponding energies are high enough to guarantee the non-degeneracy condition,  $\mu/T \ll 1$, and the ultra-relativistic limit, $\mu/T \ll 1$, too.

This allows us to calculate the energy density via Eq. (\ref{eq: energy density est})
\begin{equation}
    \label{eq: energy density est}
    \rho = \int_{0}^{\infty} E \, \,  \mathcal{F}  \, \, d^3\Vec{p},
\end{equation}
where $\mathcal{F}$ denotes the distribution function, and $\Vec{p}$ the momentum, respectively.

Taking into account the degeneracy of relativistic species, the energy density becomes
\begin{equation}
    \label{eq: Temp estadist5ica}
    \rho_{rad}(T_{est}) = g_{\ast}\frac{\pi^2}{30}T_{est}^4 \, .
\end{equation}
For the standard model, the relativistic number of degrees of freedom is $g_{\ast} = 106.75$, where all possible particles have been considered \cite{relaDoF}.

The associated temperature is known as \emph{kinetic temperature} and, for ideal gases, the kinetic and thermodynamic temperature appear equal \cite{teoriacinetica,Greiner1995}. Conversely, the temperatures are not the same for a Van der Waals gas, due to the thermodynamic interaction.

Hence, we impose the temperature/energy conditions on the radiation-dominated era,  in fulfillment with the observations coming from the cosmic microwave background (CMB) temperature.

\subsubsection{Temperature during the radiation dominated epoch}

Analogously to the pressure, evaluating at $a_{rad}$ and using the same conditions,  we obtain for the temperature
\begin{equation}
    \label{eq: temp num de part inicio radiacion}
    \left. T \right|_{a_{rad}} \approx \left. \frac{\rho V_0 a^3}{3N_{rad}c_4} \left(1 - \frac{c_1N_{inf}}{3c_4N_{rad}} \right) \right|_{a_{rad}}  \approx  \left. \frac{\rho V_0 a^3}{3N_{rad}c_4}\right|_{a_{rad}} \, .
\end{equation}

As before stated, for ideal gases, the thermodynamic and kinetic (or statistical) temperatures coincide and, so, equating Eq. (\ref{eq: Temp estadist5ica}) with Eq. (\ref{eq: temp num de part inicio radiacion}), we obtain
\begin{equation}
    \label{eq: 1/N_ic4}
    \left.\frac{1}{N_{rad}c_4}\right|_{a_{rad}} = \frac{30}{g_{\ast}\pi^2}\frac{3}{V_0a_{rad}^3T_{rad}^3} \, .
\end{equation}

On the other hand, the non-degeneracy conditions again holds, implying that the chemical potential at the radiation-dominated epoch fulfills $\mu/T \ll 1$.

So, recalling that

\begin{equation}
\frac{\mu_{rad}}{T} = - \frac{\partial S}{\partial N_{rad}},
\end{equation}
we have
\begin{equation}
    \label{eq: mu rad /T}
    \left. \frac{\mu_{rad}}{T} \right|_{a_{rad}}= \left. c_4\left[ 4-3\ln \left(\frac{\rho^3 {V_0}^4 a^{12}}{N_{rad}^4}\right) \right] \right|_{a_{rad}} \, .
\end{equation}

Let us note that, since radiation completely dominates in this epoch, then $\left.N_{rad}\right|_{a_{rad}}\approx N_i$. So, in Eq. (\ref{eq: mu rad /T}),  $N_i=\left.\rho \right|_{a_{rad}} V_0 a_{rad}^3\delta^{1/4}$ for some $\delta$ and, then, the non-degeneracy condition becomes
\begin{equation}
    \label{eq: non-degeneracy condition}
    0 \approx \left. \frac{\mu_{rad}}{T} \right|_{a_{rad}} = \left(g_{\ast}\frac{\pi^2}{30}\right)^{1/4}\frac{1}{3}\frac{1}{\delta}\left[ 4 - 3\ln(\delta)\right] \, .
\end{equation}
Solving numerically Eq. (\ref{eq: non-degeneracy condition}) for the CMB temperature escalated up to $a_{rad}$, gives us\footnote{This also sets the value for $N_i$.}
 $\delta \approx 3.79$.

Finally, using Eq. (\ref{eq: 1/N_ic4}) we get a condition for $c_4$, say
\begin{equation}
    \label{eq: c4 temp rad correcta}
    c_4 = \frac{1}{3}\left( g_{\ast}\frac{\pi^2}{30}\delta\right)^{1/4}\, .
\end{equation}

\subsubsection{Temperature  during the inflationary period}

During inflation, the energy density is constant, therefore we can set energy/temperature conditions at any value of the scale factor such that $a\in \left[a_i,a_f\right]$, where $a_i$, $a_f$ are the scale factors at the beginning and end of inflation,  respectively.

In particular, we put the conditions on the scale factor when the transition starts. Following our ansatz during the transition period, the energy density should scale as $a^{-4}$ and the CMB temperature as $a^{-1}$, providing
\begin{subequations}
\begin{align}
    \left. T \right|_{a_{trans}}  \propto \left.  T \right|_{a_{rad}} \left(\frac{a_{rad}}{a_{trans}} \right) \, ,\\
    \left. \rho \right|_{a_{trans}} \propto \left.\rho \right|_{a_{rad}} \left(\frac{a_{rad}}{a_{trans}} \right)^4 \, .
\end{align}
\end{subequations}
The proportionality constant is tuned to yield  the expected energy density at $a_{rad}$.

Consequently, Eq. (\ref{eq: Temp con num de part}) becomes
\begin{equation}
    \label{eq: temp num de part inicio inflacion}
    \left. T \right|_{a_{trans}} \approx \left. \frac{\rho V_0 a^3}{N_{i}c_1} \left(1 + \frac{N_{i}^2\alpha}{\rho {V_0}^2a^6} \right) \right|_{a_{trans}} \approx  \left. \frac{\rho V_0 a^3}{N_{i}c_1}\right|_{a_{trans}} \,,
\end{equation}
leading to a condition on  $c_1$,
\begin{equation}
    \label{eq: cond sobre c1}
    N_ic_1 \approx \left. \frac{\rho V_0 a^3}{T}\right|_{a_{trans}} \, .
\end{equation}

\subsubsection{Temperature during the matter-dominated epoch}

In the matter-dominated phase,
Eq. (\ref{eq: Temp con num de part}) acquires the form
\begin{equation}
    \label{eq: temp num de part inicio materia}
    \left. T \right|_{a_m}  \approx  \left. \frac{\rho V_0 a^3}{N_{dust}c_5}\right|_{a_{m}} \,,
\end{equation}
where $a_m$ indicates the time of matter domination.

For simplicity, we can set $a_m = a_{eq}$, where $a_{eq}$ is the scale factor at which the equivalence occurs. Remembering that the energy density scales as $a^{-4}$ for radiation and $a^{-3}$ for matter  while the temperature scales as $a^{-1}$,  the condition on the constant $c_5$ reduces to
\begin{equation}
    \label{eq: cond sobre c5}
    c_5 \approx \left. \left( \frac{V_0}{N_i} \frac{\rho}{T}\right)\right|_{a_{rad}} a_{eq} \, .
\end{equation}

\subsubsection{The parameter $\zeta$}

To obtain the value for the parameter $\zeta$, we make use of the temperature equation of state, imposing that the  temperature of the (remnant) inflationary fluid is the same as that of the CMB, implying
\begin{equation}
    1 = T_{CMB,0}  \left. \left(\frac{\rho_{inf} V}{c_1 N_{inf}} \right|_{0} \right)^{-1}.
\end{equation}
Setting $T_{CMB,0} = 2.7 K$, $\left. \rho_{inf} \right|_0 = \rho_{crit,0} +\Omega_{\Lambda,0}$, and $\left.N_{inf}\right|_0 = \zeta N_i$ we get:

\begin{equation}
    \zeta_{max} = T_{CMB,0} \frac{c_1N_i}{\rho_{crit,0} \Omega_{\Lambda,0} V_0} \approx 8.9 \times 10^{-2} .
\end{equation}

The above value represents the maximum value the constant $\zeta$ can have, to get the present universe's acceleration. If we assume a lower value, the cosmological constant is not the only source of late-time acceleration, but rather a component of it. Hence, the density parameter related to the cosmological constant is
\begin{equation}
    \rho_{\Lambda} + \left.\rho_{inf} \right|_{a_0}= \rho_{crit,0}\left(1-\sum_i \Omega_{i,0}\right) ,
\end{equation}
where we set the spatial curvature equal to zero, and $i$ denotes the species (dark matter, baryonic matter, radiation, etc). This ultimately leads to a change in the value of the constant since

\begin{equation}
    \Lambda = 8\pi G \left[\rho_{crit,0}\left(1-\sum_i \Omega_{i,0}\right) - \frac{c_1N_i\zeta T_0}{V_0}\right] .
\end{equation}

So for $\zeta = \zeta_{max}$, $\Lambda = 0$.

\subsection{Solving the continuity equation}

With the above values for the parameters, we obtain a consistent model that yields first a constant energy density and then  an energy density that scales as $a^{-4}$ at the required time.  Also, the (numerically) calculated number of e-foldings matches the previous ansatz.

\begin{figure}[!htb]
    \centering
    \includegraphics[width=240pt]{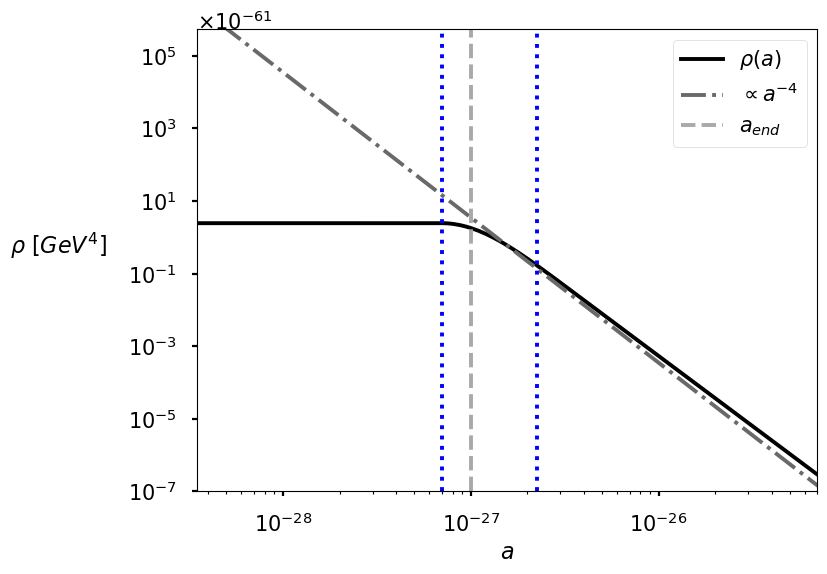}
    \caption{Energy density during the transition/decaying process in logarithm scale (continuous line). The vertical dotted lines denote the region where the transition/decaying process occurs. The vertical dashed line denotes the scale factor when inflation ends / radiation begins (i.e. $a_{end}$), and the dotted-dashed line represents an energy density that is proportional to $a^{-4}$.}
    \label{fig: densidad de energia transicion final log}
\end{figure}

The numerical solution exhibits three different behaviors for the energy density. When the inflationary fluid dominates, the energy density is  constant (left part of the slope); when radiation dominates, the energy density (in logarithm scale) becomes a straight line with a slope of $-4$ (See Fig. \ref{fig: densidad de energia transicion final log}).
Finally, between these two different behaviors, there exists a transition period.

\begin{figure}[!htb]
    \centering
    \includegraphics[width=240pt]{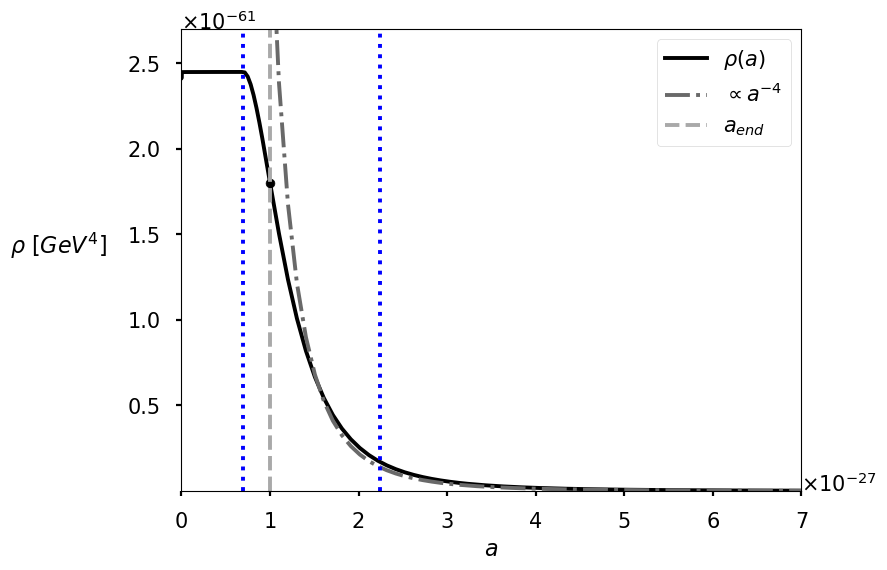}
    \caption{Energy density during the transition/decaying process (continuous line). The vertical dotted lines denote the region where the transition/decaying process occurs.  The vertical dashed line denotes the scale factor when inflation ends / radiation begins (i.e. $a_{end}$), and the dotted dashed line represents an energy density that is proportional to $a^{-4}$.}
    \label{fig: densidad de energia transicion}
\end{figure}

In this respect, Fig. \ref{fig: densidad de energia transicion} shows the transition period between inflation and radiation. As we can see, there is a change in the concavity of the plot. The inflection point is in fact the point where the dominating fluid changes and, clearly, it also denotes the end of inflation/beginning of radiation.

\subsubsection{End of inflation}

Obtaining the inflection point can be computed by consider the slow roll condition the Hubble parameter. Inflation ends as $\left.\rho + 3P \right|_{a_f} = 0$, for some $a_{f}$, which denotes the scale factor when inflation ends.

With these two above well-known conditions, we can compute the pressure $P$ from the numerical solution for  $\rho + 3P$, where the change of sign determines $a_f$. Hence, the scale factors at which the sign of $\rho + 3P$ changes are: $a =  10^{-27}$ and $a = 1.1 \times 10^{-27}$, i.e., $a_f \in \left[10^{-27}, 1.1\cdot 10^{-27}\right]$. Accordingly, for simplicity, we could take  $a_f = (1+1.1)/2 \cdot 10^{-27} = 1.05 \cdot 10^{-27}$.

\subsubsection{End of the transition/decaying process}

We can here finally calculate the moment when the transition ends, i.e., when the condition $\left. 3N_{rad}c_4 \right|_{a^{\ast}}  \gg \left. N_{inf}c_1 \right|_{a^{\ast}} $ breaks down, namely

\begin{equation}
\left. \frac{c_1N_{inf}}{3c_4N_{rad}} \right|_{a^{\ast}}  \not\ll 1\,.
\end{equation}

Our numerical computation yields that for $a^{\ast}=3.2a_{trans} \approx 2.4 \times 10^{-27}$, the condition truly breaks as we have

\begin{equation}
\left. \frac{c_1N_{inf}}{3c_4N_{rad}} \right|_{2.4a_{trans}}  \approx 0.1\,.
\end{equation}

So, the transition starts at
\begin{equation}
a_{trans} = 7\times 10^{-28},
\end{equation}
and ends at\footnote{Note that inflation and the transition epoch do not end simultaneously.}
\begin{equation}
a=3.2a_{trans} \approx 2.24 \times 10^{-27}.
\end{equation}

Last but not least, we here remark  that we do not consider any mechanism to create and destroy particles since we presume baryogenesis to occur before the existence of our fluid and then we handle the densities of particles through $n$ as reported above.

\subsubsection{Temperature behavior}

\begin{figure}[!htb]
    \centering
    \includegraphics[width=240pt]{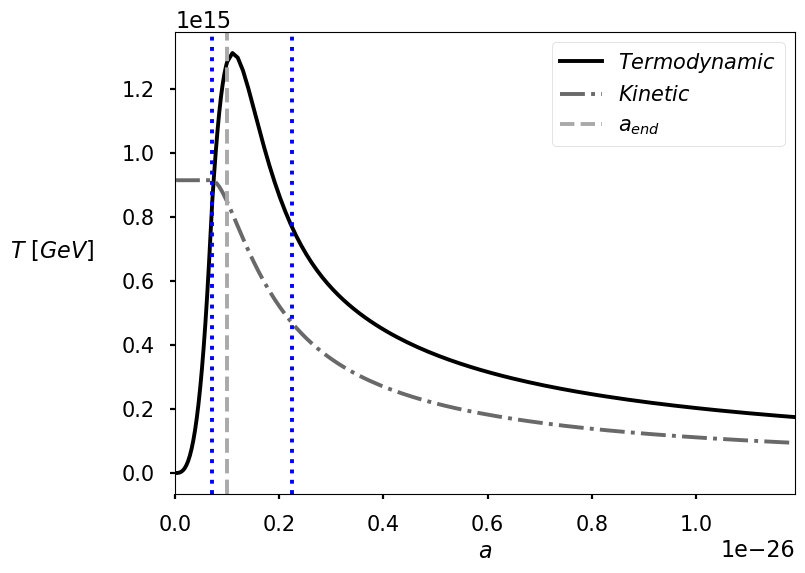}
    \caption{Thermodynamic (continuous line) and Statistic/kinetic temperature (dotted-dashed line) with the non-degeneracy condition full-filled for $\zeta = 0.087$ and $\lambda = 0.8$ (which implies that the radiation fluid represents 65\% of the total). The vertical dotted lines denote the region where the transition/decaying process occurs. The vertical dashed line denotes the scale factor when inflation ends / radiation begins (i.e. $a_{end}$).}
    \label{fig: both temperatures correct}
\end{figure}

Remarkably, we can here spend some comments on the underlying temperature. In Fig. \ref{fig: both temperatures correct}, we plot the  thermodynamic and kinetic temperatures and note that, albeit  radiation dominates, the temperatures are not equal at $a=a_{rad}$.

This is due to the existence of other fluids, matter and inflationary constituents, that impose such a deviation.

However, for a total decay,  i.e., assuming $N_{inf}$ decaying totally into $N_{rad}$, the corresponding plots match, as drawn in Fig. \ref{fig: both temperatures}.
Nevertheless, both temperatures tend to the CMB temperature as the scale factor increases, namely as the universe expands.

Conversely, as we approach the beginning of inflation, we can observe that the thermodynamic temperature  tends to zero. This does not imply that particles do not have kinetic energy, but rather that the thermodynamic interaction dominates over the kinetic part. Recalling that the constants $c_1$ and $c_2$ are also the heat capacities of the inflationary fluid, the  condition $c_2 <0$ means that,  when the energy of the system decreases (at constant pressure), its temperature increases.

In other words, the thermodynamic interaction contribution to the (thermodynamic) temperature at the beginning of inflation is large enough to counter its kinetic counterpart, albeit,  as the universe expands, the thermodynamic interaction term washes off, the fluid becomes ideal, and the temperatures tend to coincide.

\section{Healing Big Bang inconsistencies through thermodynamic inflation}\label{sezione4}

Our model assumes the existence of a fluid whose properties mimes those of an inflaton field, but being quite different than it. Our treatment, however, does not directly face the main issues related to the standard Big Bang paradigm that, in general, are faced by standard inflation, through the use of a scalar field.

In what follows, it appears essential, then, to investigate the flatness and horizon problems to check the suitability of our thermodynamic inflation.

\subsection{The flatness problem}

Throughout the manuscript, it was assumed - for simplicity - that the spatial curvature, $k$, is \emph{exactly} zero.

This assumption contradicts the basic demands of the standard Big Bang paradigm that requires to explain the reasons behind a so small spatial curvature. As above stated, the inflationary mechanism appears essential, being capable of explaining how the universe becomes so flat, avoiding \emph{de facto} a fine-tuning problem on $k$.

Hence, we can first notice that the inflationary mechanism is responsible for fixing $k$ to small values, as expected. Second, let us  notice that the continuity equation in Eq. (\ref{eq: continuity}) does not include explicitly the spatial curvature term itself. Thus, our fluid, capable of speeding the universe up, can be reconsider as $k\neq0$ without losing any generality and its main properties might remain almost unaltered.

\begin{figure}[!htb]
    \centering
    \includegraphics[width=240pt]{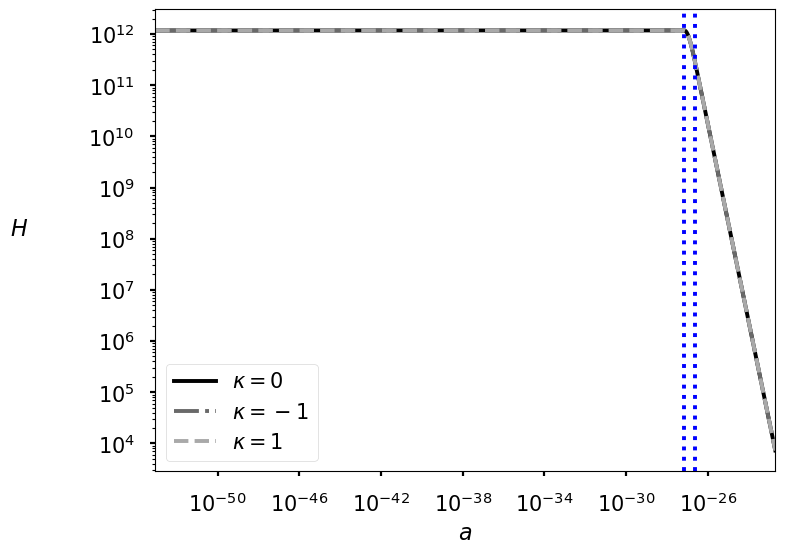}
    \caption{Hubble function $H$ for $\kappa = -1,0,1$ (dotted-dashed, continuous and dashed respectively). The horizontal axis (representing the scale factor $a$) with a logarithm scale. The vertical dotted lines denote the region where the transition/decaying processes hold.}
    \label{fig: H with curvature}
\end{figure}

\begin{figure}[!htb]
    \centering
    \includegraphics[width=240pt]{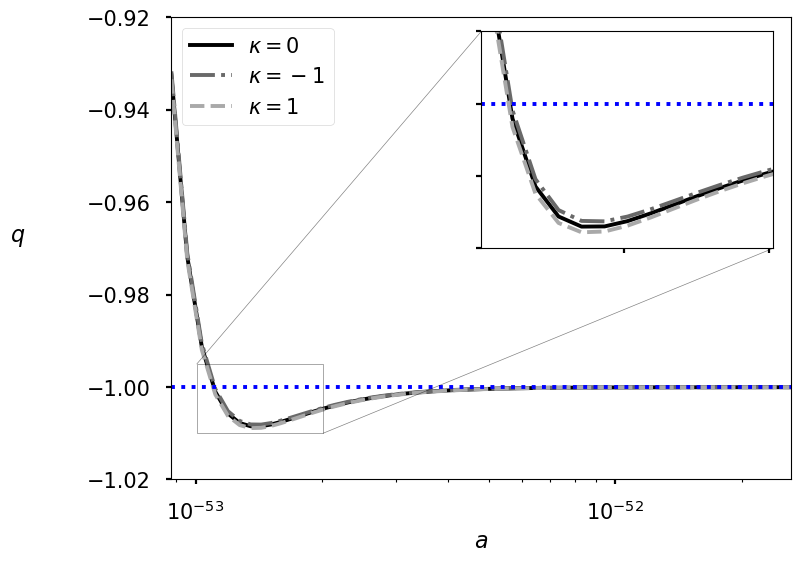}
    \caption{Deceleration parameter $q$ for $\kappa = -1,0,1$ (dotted-dashed, continuous and dashed respectively). The horizontal axis (representing the scale factor $a$) with a logarithm scale. The vertical dotted lines denote the region where the transition/decaying processes hold.}
    \label{fig: q with curvature}
\end{figure}

The acceleration mechanism can be displayed, regardless the spatial curvature, analyzing the deceleration parameter, $ q = - \Ddot{a}a/\Dot{a}^2$, as well as the Hubble parameter, as shown in Figs. \ref{fig: H with curvature} and \ref{fig: q with curvature}, where we used $k=\kappa/r_0^2$, with $\kappa = \left\{-1,0,1\right\}$ and $r_0 = 3.21_{-1.01}^{+4.78} \times 10^{42} GeV^{-1}$ \cite{curvradius}.

Nevertheless, taking as above three intervals: $(a_i, a_f)$, $(a_f, a_{eq})$, $(a_{eq},a_0)$, and calculating the density parameter, we obtain

\begin{equation}
    \frac{\Omega_i^{-1} - 1}{\Omega_0^{-1} - 1} = \frac{\rho_0a_0^2}{\rho_{eq}a_{eq}^2} \frac{\rho_{eq}a_{eq}^2}{\rho_fa_f^2}\frac{\rho_fa_f^2}{\rho_{i}a_{i}^2},
\end{equation}

implying the following condition for the density parameter,

\begin{equation}
\begin{split}
    \left| 1 - \Omega_0^{-1} \right|  = & \left| 1 -\Omega_i^{-1} \right| \exp \left[- \left(N  + \frac{1}{2}\ln\left(z_{eq} \right) + \ln\left(\frac{T_f}{T_{eq}}\right) \right. \right.\\
    & - \left. \left. \ln \left(\frac{\rho_f}{\rho_i}\right) \right) \right]    \\
     \approx & \left| 1 -\Omega_i^{-1} \right| 3.12\times 10^{-52},
\end{split}
\end{equation}
that solves the issue related to flatness, switching to the use of a fluid, rather than an inflaton field.

\subsection{The horizon problem}

In analogy to the flatness problem, to solve the horizon problem, another caveat of the standard Big Bang paradigm, we need to show that

\begin{equation}
    \label{eq: horizon constriction}
    \left. r_c \right|_{i} = \left. aH \right|_{i} \ge \left. r_c \right|_{0} =  \left. aH \right|_{0} \, .
\end{equation}

Accordingly, we may consider three intervals along the universe's history,  $(a_i, a_f)$, $(a_f, a_{eq})$, $(a_{eq},a_0)$, corresponding to inflation, radiation, and matter dominated epochs, respectively \cite{PeterColes}.

From Eq. (\ref{eq: horizon constriction}), we have

\begin{equation}
    \frac{H_ia_i}{H_fa_f} \le \frac{H_0a_0}{H_fa_f} = \frac{H_0a_0}{H_{eq}a_{eq}}\frac{H_{eq}a_{eq}}{H_fa_f}.
\end{equation}

Now, since $\frac{H_0a_0}{H_{eq}a_{eq}} = \left(\frac{a_0}{a_{eq}}\right)^{-1/2}$, $\frac{H_{eq}a_{eq}}{H_fa_f} = \left(\frac{a_{eq}}{a_f}\right)^{-1}$; $a_0/a_{eq} \approx z_{eq}$ and $T \propto a^{-1}$ we have:

\begin{equation}
\begin{split}
    N  =  \ln \left(\frac{a_f}{a_i}\right) \ge & \frac{1}{2} \ln (z_{eq}) + \ln\left(\frac{T_f}{T_{eq}}\right) - \ln \left(\frac{H_f}{H_i}\right) \, . \\
      \ge &  58.59
\end{split}
\end{equation}

On the other hand, as explained in Sec. \ref{sezione3}, we can interpret our thermodynamic phase transition as the process that gives rise to the inflationary epoch, meaning that the scale factor at which inflation started must be (at most) equal to the scale factor at which the phase transition happened.

Following the same procedure, we get that the phase transition for this model occurs at $a \approx 2.37 \times 10^{-58}$, since $a_f \approx 10^{-27}$, we get that the maximum number of e-foldings is $70.51$.

Hence the number of e-foldings is bounded as follows,

\begin{equation}
N_e \in (58.6,70.5)\,,
\end{equation}
perfectly compatible with the expectations made in standard inflationary scenarios.

Accordingly, our treatment appears quite similar in explaining the inflationary mechanism, albeit very different in using a fluid rather than a field for accelerating the universe at primordial-times.

Even if the model appears promising, we can enumerate and compare the main good characteristics with pending points that seem to deserve further investigation, as follows.

\begin{itemize}
    \item[-] The model acts to accelerate the universe without scalar fields, but requires a further exotic form of matter with pressure, whose nature is identified through a fundamental equation that has to be formulated \emph{a priori}.
    \item[-] The model reproduces the inflationary main features, such as e-foldings, $a_i$, $a_f$, etc., solving  the flatness and horizon problems,  requiring however particles to form, but may appear fine-tuned in choosing the free constants.
    \item[-] The acceleration is induced by a standard phase transition and provides a natural ends, induced by particle production, but the existence of small perturbations cannot immediately be  accounted inside our theoretical scheme and, so, it clearly deserves future investigation.
\end{itemize}

\section{Outlooks and perspectives} \label{sezione5}

We investigated a cosmic thermodynamic fluid acting to accelerate the universe at primordial-times, mimicking the inflationary period without the need of an inflaton field. To construct the fluid, we invoked GTD as a   foundational background and we illustrated the conditions under which inflation can arise using this fluid.

To do so, we proposed a fundamental equation, whose functional form may appear particularly complicated as it has to dominate over radiation and matter,  throughout the inflationary phase.

In this respect, we also included standard matter, under the form of dust, plus radiation and other constituents, in order to characterize the universe at primordial-times.

To do so, including unspecified real fluids into the constitutive equation, we demonstrated that it is possible to recover an inflationary end similar to the standard picture. Moreover,  the corresponding dynamics recovers the number of e-foldings, needful to accelerate the universe, as well as suitable initial and final stages for the inflationary epoch.

Hence, conceptualizing the universe as an overall thermodynamic system, by  selecting the free constants of our model to provide a rapid expansion, we demonstrated that a quasi-de Sitter phase can be naively reached, mimicking inflation.

Physical consequences on the choices of the underlying constants have been carefully discussed, as well as effects on pressure, free energy, density and temperatures, explored to guarantee that our outcomes match with cosmological constraints.

Precisely, we attributed the onset of the cosmic speed up to a phase transition, invoking that our fluid constituent is not barotropic, but rather appearing as a real fluid.

Subsequently, we investigated how inflation ends within this context and established that even the late-time accelerated expansion of the universe can easily be recovered, unifying \emph{de facto} the process of inflation with dark energy.

To do so, we clearly compared our overarching GTD fluid model with the most common flatness and horizon problems, providing a theoretical discussion toward its use in alternative to scalar fields.

Even though promising, our approach suffers from the determination of the fundamental nature of cosmic fluids inserted into the thermodynamics associated with GTD. Consequently, future works will explore how to clarify the nature of fundamental constituents behind the choice of our equations of state. Moreover, the caveat of passing through a thermodynamic domain, offered by GTD, to a geometric one, directly related to the FRW spacetime, will also be featured. Further studies could also shed light on multiple phase transitions in the contexts of extended multi-fluid approach related to our scheme. Finally, the role of small perturbations might be clarified and also consequences in the primordial power spectrum.

\section*{Acknowledgments}
OL expresses his gratitude for the hospitality of the Instituto de Ciencias Nucleares at UNAM University during the period in which this paper was conceived. He is also grateful to Roberto Giambo' and Marco Muccino for their useful discussions on the topic of this paper.

\newpage
\onecolumngrid
\appendix

\section{Temperature and dynamics}

In this appendix, we report additional plots related to the temperature and deceleration parameter of our model.

\begin{figure}[!htb]
    \centering
    \includegraphics[width=240pt]{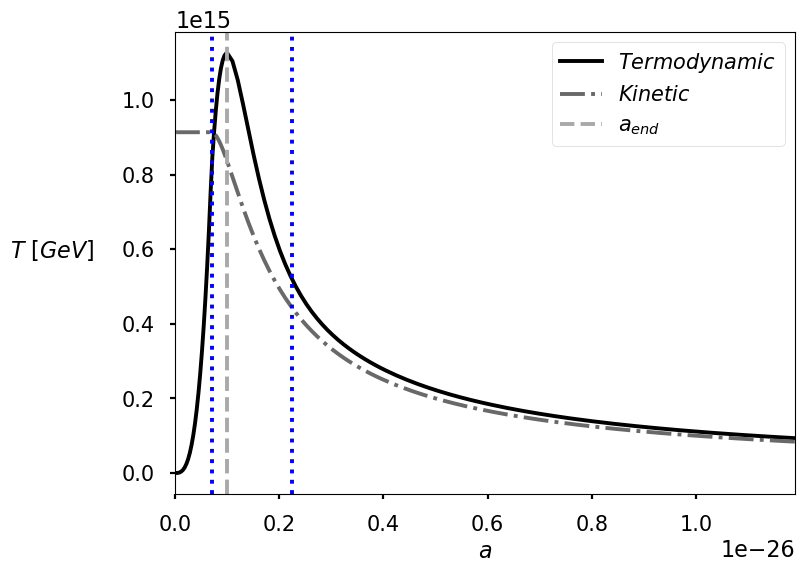}
    \caption{Thermodynamic (continuous line) and Statistic/kinetic temperature (dotted-dashed line) with the non-degeneracy condition full-filled for $\zeta = 0$, $\lambda = 0.9$ (which implies that the radiation fluid represents 90\% of the total). The vertical dotted lines denote the region where the transition/decaying process is being held. The vertical dashed line denotes the scale factor when inflation ends / radiation begins (i.e. $a_{end}$).}
    \label{fig: both temperatures}
\end{figure}

\begin{figure}[!htb]
    \centering
    \includegraphics[width=240pt]{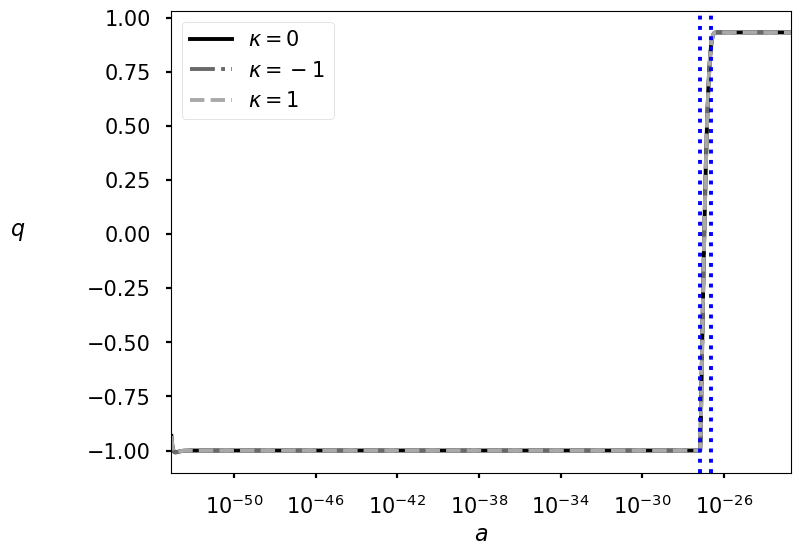}
    \caption{Deceleration parameter $q$ for $\kappa = -1,0,1$ (dotted-dashed, continuous and dashed respectively). The horizontal axis (representing the scale factor $a$) with a logarithm scale. The vertical dotted lines denote the region where the transition/decaying processes hold.}
    \label{fig: q with curvature total}
\end{figure}

\end{document}